\documentclass[prl,aps,twocolumn,superscriptaddress,longbibliography,notitlepage]{revtex4-1}
\usepackage[colorlinks,linkcolor=darkblue,citecolor=darkblue,urlcolor=darkblue]{hyperref}
\usepackage{amsmath}
\usepackage{amsfonts}
\usepackage{xcolor}
\usepackage{verbatim}
\usepackage{graphicx}
\usepackage{esint}
\usepackage{float}
\usepackage{amssymb}
\usepackage{mathrsfs}
\usepackage{bm}
\usepackage{bbm}
\definecolor{darkblue}{rgb}{0,0,0.6}
\usepackage{physics}
\usepackage{mathtools}
\newcommand{\beq}{\begin{equation}}
\newcommand{\eeq}{\end{equation}}

\newcommand*\de{\mathop{}\!\mathrm{d}}
\usepackage{soul}

\begin{document}

\title{Two-dimensional crystals far from equilibrium}

\author{Leonardo Galliano}

\affiliation{Laboratoire Charles Coulomb (L2C), Universit\'e de Montpellier, CNRS, 34095 Montpellier, France}

\author{Michael E. Cates}

\affiliation{Department of Applied Mathematics and Theoretical Physics, University of Cambridge, Wilberforce Road, Cambridge CB3 0WA, United Kingdom}

\author{Ludovic Berthier}

\affiliation{Laboratoire Charles Coulomb (L2C), Universit\'e de Montpellier, CNRS, 34095 Montpellier, France}

\affiliation{Yusuf Hamied Department of Chemistry, University of Cambridge, Lensfield Road, Cambridge CB2 1EW, United Kingdom}

\date{\today}

\begin{abstract}
  When driven by nonequilibrium fluctuations, particle systems may display phase transitions and physical behaviour with no equilibrium counterpart. We study a two-dimensional particle model initially proposed to describe driven non-Brownian suspensions undergoing  nonequilibrium absorbing phase transitions. We show that when the transition occurs at large density, the dynamics produces long-range crystalline order. In the ordered phase, long-range translational order is observed because equipartition of energy is lacking, phonons are suppressed, and density fluctuations are hyperuniform. Our study offers an explicit microscopic model where nonequilibrium violations of the Mermin-Wagner theorem stabilize crystalline order in two dimensions. 
\end{abstract}

\maketitle

The physical world can be described using systematic theoretical tools and general principles governing the existence and physical behaviour of various phases of matter and of the transitions between them~\cite{tabor1991gases}. Equilibrium statistical physics is an elegant theoretical construction based on minimal assumptions from which phase transitions can be predicted~\cite{callen1998thermodynamics,landau2013statistical}. Equilibrium and symmetry principles constrain the range of possible phases that particle systems exhibit~\cite{chaikin1995principles}. At large density, an assembly of repulsive particles crystallises in dimensions $d>2$, but equilibrium thermal fluctuations prevent the existence of long-range translational order in $d \leq 2$~\cite{mermin1968crystalline}. This is a consequence of the Mermin-Wagner theorem~\cite{mermin1966absence}. As a result, $2d$ equilibrium systems only exhibit quasi-long-range translational order, which generically occurs passing through an intermediate hexatic phase with quasi-long-range orientational order~\cite{kosterlitz1973ordering,halperin1978theory,young1979melting}. In $1d$, the system is disordered at any finite temperature. 

Equilibrium statistical mechanics is often insufficient to describe physical phenomena and systems that evolve far from equilibrium: active matter is not driven by thermal fluctuations~\cite{marchetti2013hydrodynamics}, fluids can be stirred by shear flows~\cite{larson1999structure}, biological systems are alive~\cite{nelson2004biological}, and granular systems are dissipative~\cite{jaeger1996granular}. The $2d$ ordering transition of colloidal~\cite{zahn1999two,thorneywork2017two}, granular~\cite{reis2006crystallisation,olafsen2005twodimensional,komatsu2015roles}, and active~\cite{digregorio2018full,klamser2018thermodynamic,caprini2020hidden,briand2016crystallisation,baconnier2022selective} materials has been studied. While the nature and sequence of phase transitions may be affected by nonequilibrium effects, the absence of long-range translational order has thus far been robustly confirmed~\footnote{A numerical study~\cite{gowrishankar2016nonequilibrium} reports an ordered phase of vortices in an active polar fluid, but the maximum linear size of the lattice is five lattice units, making a discussion of the thermodynamic limit difficult.}. This situation contrasts with orientational order: flocking active matter is a celebrated example where nonequilibrium fluctuations are strong enough to produce violations of the Mermin-Wagner theorem leading to long-range magnetic order in driven $2d$ XY models~\cite{toner1995longrange}. A similar result was shown to hold for sheared $2d$ $O(2)$ models~\cite{nakano2021long}.  

Absorbing phase transitions represent a broad class of nonequilibrium transitions with many physical applications~\cite{henkel2008non}. Corresponding microscopic models are defined from local dynamic rules which do not rely on equilibrium assumptions and break detailed balance. Driven colloidal suspensions~\cite{pine2005chaos} have been described using random organisation particle models~\cite{corte2008random}, which display an absorbing phase transition presumably lying in the conserved directed percolation (CDP) universality class~\cite{menon2009universality,tjhung2016criticality}. Upon increasing the density the system transitions from an absorbing to a disordered diffusing phase whose non-thermal behaviour has been explicitly demonstrated~\cite{schrenk2015evidence}. At the critical point, the system becomes hyperuniform~\cite{hexner2015hyperuniformity,tjhung2015hyperuniform,weijs2015emergent,wilken2020hyperuniform,torquato2018hyperuniform}, and hyperuniformity extends to the active disordered phase if the center of mass is conserved during binary collisions~\cite{hexner2017noise,lei2019hydrodynamics}. This model was also recently studied at large density in relation to random close packing~\cite{wilken2021random,ness2020absorbing,wilken2022random}, while a recent granular experiment suggests that coupling to structural ordering can even change the universality class of the transition~\cite{ghosh2022coupled}.  

Here we analyse the interplay between nonequilibrium random organisation dynamics and structural order at large density in a $2d$ isotropic version of the model~\cite{tjhung2015hyperuniform} with center-of-mass conservation~\cite{hexner2017noise}. The system orders as density increases but the nature of this transition differs from both equilibrium and known nonequilibrium situations. Remarkably, since equipartition of energy is lacking, long-wavelength phonons are suppressed and long-range translational order can emerge. As a result, hyperuniform two-dimensional crystals spontaneously form far from equilibrium.

\begin{figure}
   \includegraphics[width=0.48\textwidth]{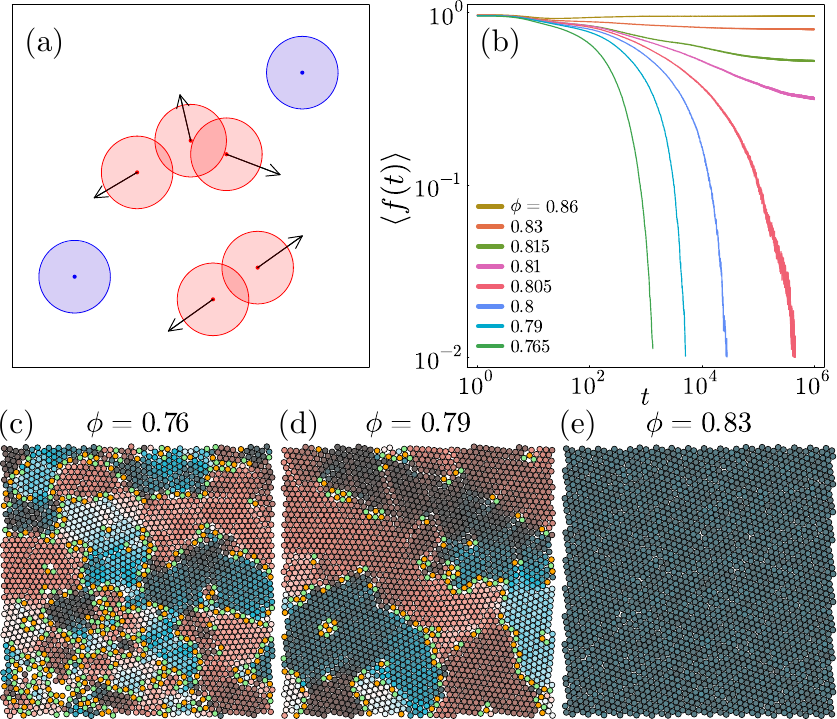}
    \caption{{\bf Model and absorbing phase transition.} (a) Dynamic rules: at each time step $t$, pairs of overlapping particles (red) are displaced in opposite directions by the same random amount. (b) Time dependence of the average fraction of active particles starting from random initial conditions for $N = 4 \times 10^4$. The activity vanishes for low $\phi$, whereas a dynamic steady state with a finite activity is reached at large $\phi$. (c-d) Absorbing configurations resemble polycrystals (the colour codes for orientational order), while (e) active configurations are fully ordered (linear size in (c-e) is $L=50$).}
    \label{fig:model}
\end{figure}

We study a two-dimensional off-lattice model of $N$ particles of diameter $\sigma$ in a square simulation box of linear size $L$ with periodic boundary conditions. The packing fraction $\phi= (N \pi \sigma^2) / (4L^2)$ is the first control parameter. At each discrete time $t$, pairs of overlapping particles are displaced in opposite directions along the axis connecting their centers, see Fig.~\ref{fig:model}(a), so that particles move away from each other by a random amount drawn from a flat distribution in the interval $[0,\epsilon]$; $\epsilon$ is the second control parameter. If a particle overlaps with several neighbours, its displacement is the sum of each pairwise contribution. These microscopic rules conserve the position of the center of mass~\cite{hexner2015hyperuniformity}. Particles that jump at time $t$ are active. The global activity, $f(t)$, is the fraction of active particles at time $t$. It is the order parameter for the absorbing phase transition. We studied $\epsilon = 0.2$, $0.1$ to push the critical density to a large value~\cite{tjhung2016criticality,wilken2021random}, and have varied the remaining parameter $\phi$, finding equivalent results. 

We first follow the evolution of the system starting from fully random initial conditions for $\epsilon=0.1$ and several packing fractions. As shown in Fig.~\ref{fig:model}(b), the evolution of $\langle f(t) \rangle$ reveals two distinct types of behaviours. The system reaches an absorbing phase with vanishing activity at low $\phi$, while it remains active over the numerical time window at larger $\phi$. Visual inspection of the particle configurations during these dynamics reveals that the system develops structurally ordered domains which coarsen with time. At low $\phi$, the coarsening is interrupted after a finite time, and absorbing configurations resemble polycrystals, see Figs.~\ref{fig:model}(c-d). In finite systems at large density, a single crystalline domain eventually fills the box while keeping a finite level of steady state activity, see Fig.~\ref{fig:model}(e). As usual for coarsening~\cite{bray2002theory}, the timescale to reach such ordered steady states diverges with system size.

\begin{figure}
   \includegraphics[width=0.5\textwidth]{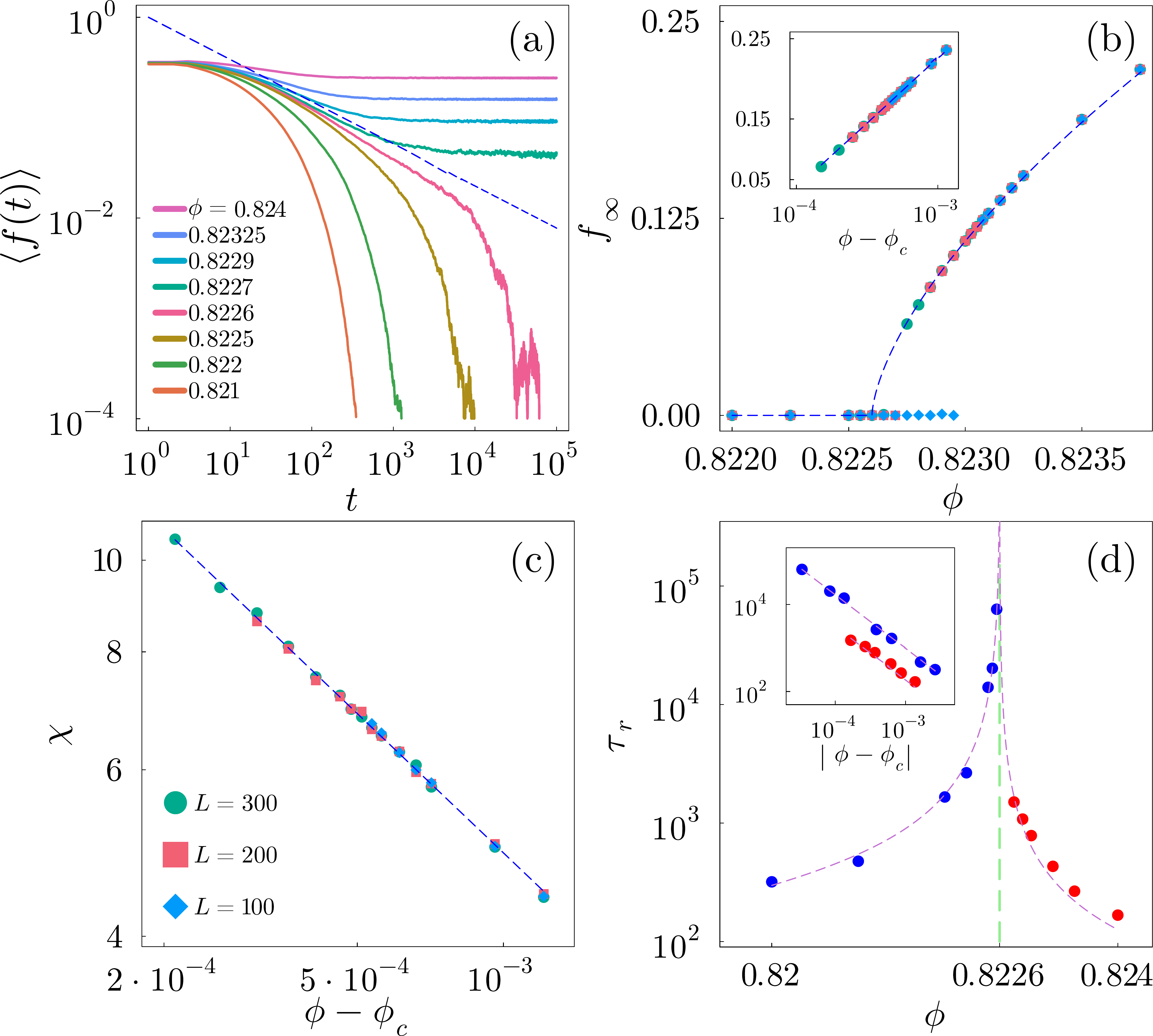}
  \caption{{\bf Conserved directed percolation universality class.} 
  (a) The time dependence of the activity starting from a fully ordered structure reveals a critical packing fraction $\phi_c$ where $\langle f(t) \rangle \sim t^{-\alpha}$ shown with dashed line for $\alpha=0.42$, $N=10^5$. (b) Critical scaling of the average activity for three system sizes. The dashed line is Eq.~(\ref{eq:beta}) with $\beta=0.64$, shown in logscale in the inset. (c) Diverging fluctuations of the activity at steady state. The dashed line is Eq.~(\ref{eq:gamma}) with $\gamma=0.49$. (d) Critical scaling of the relaxation time starting from an ordered configuration in the absorbing (blue) and active (red) phases. The dashed line is Eq.~(\ref{eq:nu}) with $\nu_\parallel=1.3$ shown in logscale in the inset. In all panels, $\phi_c = 0.8226$.}
  \label{fig:dyncriticality}
\end{figure}

Previous work~\cite{corte2008random,menon2009universality,tjhung2016criticality} suggested that random organisation models belong to the conserved directed percolation (CDP) universality class. Numerical results are consistent with this hypothesis, although the set of critical exponents are so close to the DP (directed percolation) class that the distinction between the two is numerically challenging. We now establish that the exponents we measure are consistent both with CDP and previous numerics. To analyse steady state physics above the transition we randomly select a fully ordered configuration reached after coarsening has finished well above $\phi_c$, and follow its relaxation towards steady state at different $\phi$, as shown in Fig.~\ref{fig:dyncriticality}(a). In practice we fix $N=95706$ and sligthly vary the system size $L \in [302.03, 302.58]$ to change $\phi$. This protocol allows us to measure the evolution of the characteristic timescale $\tau_r$ to reach an absorbing state (below $\phi_c$) or active steady state (above $\phi_c$), while the relaxation of the activity appears algebraic at $\phi_c$, $\langle f(t) \rangle \sim t^{-\alpha}$. In Fig.~\ref{fig:dyncriticality}(a), we have used the known CDP value, $\alpha = 0.42$.

\begin{figure*}
  \includegraphics[width=0.99\textwidth]{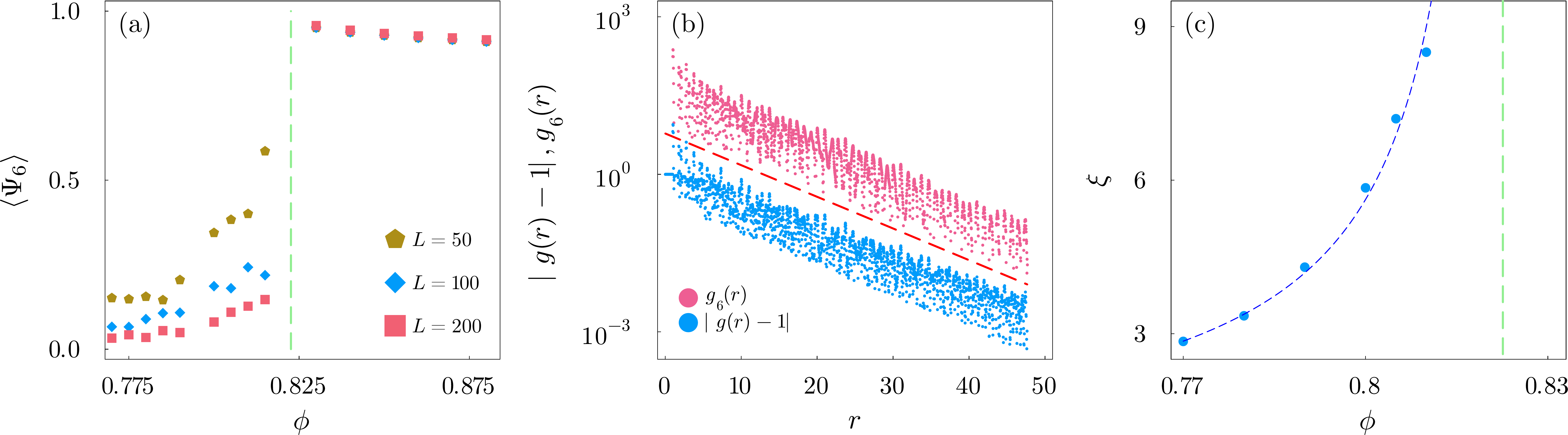}
  \caption{{\bf Transition to long-range order.}  (a) Discontinous jump of the global bond-orientational order parameter for different system sizes. The vertical  dashed line indicates $\phi_c$. (b) Coupled decay of the pair correlation function (blue) and bond-orientational correlation function (pink, shifted vertically for clarity) for $\phi=0.805$ and $L=200$. Dashed is a fit to exponential decay, $\sim e^{-r/\xi}$. (c) Diverging correlation length $\xi$ quantifying the typical size of crystalline domains. The blue dashed line is $\xi \sim (\phi - \phi_c)^{-\nu_{\perp}}$ with $\nu_\perp = 0.8$ known from CDP.}
    \label{fig:ordering}
\end{figure*}

Above $\phi_c$, we expect the steady state activity to obey 
\begin{equation}
    f_\infty \equiv \langle f(t \to \infty) \rangle \sim (\phi-\phi_c)^{\beta}, 
    \label{eq:beta}
\end{equation}
with $\beta = 0.64$ for CDP. The corresponding numerical results are shown in Fig.~\ref{fig:dyncriticality}(b) for several system sizes. They follow Eq.~(\ref{eq:beta}). We define a susceptibility associated to the fluctuations of the activity in steady state, $\chi = N (\expval{f^2}- f_\infty^2 )$, which is expected to diverge as
\begin{equation}
    \chi \sim (\phi-\phi_c)^{-\gamma}, 
    \label{eq:gamma}
\end{equation}
where $\gamma = 0.49$ was reported~\cite{tjhung2016criticality}. The agreement with the data in Fig.~\ref{fig:dyncriticality}(c) is excellent. We can finally measure relaxation times on both sides of the transition. For $\phi<\phi_c$ we define the relaxation time $\tau_r$ as the number of steps needed to reach an absorbing state with $f(t)=0$. Above $\phi_c$, it is defined as the first time at which the value of $\expval{f(t)}$ attains its stationary value $f_\infty$, within a tolerance which we arbitrarily set to 0.005. In Fig.~\ref{fig:dyncriticality}(d), these two timescales are shown to diverge at the critical point as 
\begin{equation}
     \tau_r\sim\abs{\phi-\phi_c}^{-\nu_{\parallel}},
     \label{eq:nu}
\end{equation}
with an exponent $\nu_{\parallel}=1.3$ again consistent with the CDP value. Overall, our numerics demonstrates that despite the much larger value of the critical packing fraction and observed structural order, the dynamic criticality is compatible with all studies performed previously in the absence of structural order. In this respect our conclusions disagree with the recent experimental study~\cite{ghosh2022coupled}.

\begin{figure*}
   \includegraphics[width=\textwidth]{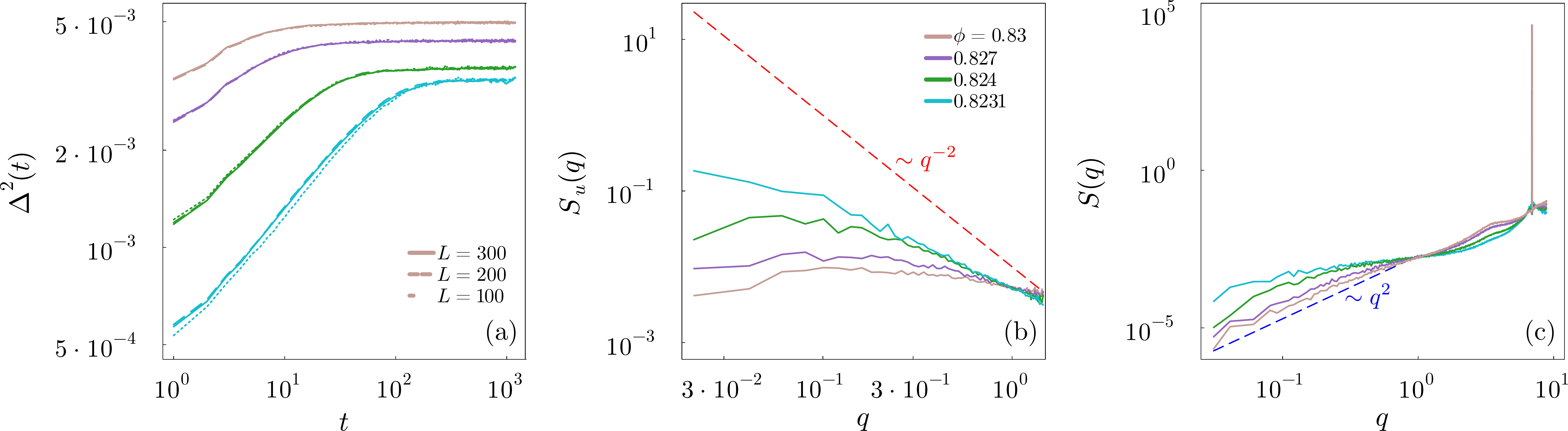}
  \caption{{\bf Long-range translational order and hyperuniformity.} (a) Mean-squared displacement at several packing fractions for three system sizes above $\phi_c$. The plateau values do not depend on $L$, and crystal are thus stable. The plateau values increase slightly with $\phi$, since multiple overlaps become more common at larger $\phi$. (b) Structure factor of the displacement field for several packing fractions at $L=300$. The red dashed line represents the thermal limit where equipartition holds with $S_{u} (q) \sim 1/q^{2}$. (c) Static structure factor $S(q)$ for several packing fractions at $L=200$. It vanishes as $S(q) \sim q^2$ at low $q$ (shown as dashed line) in the active phase: $2d$ active crystals are hyperuniform.}
\label{fig:activecrystal}
\end{figure*}

What distinguishes our study from previous ones, however, is that structural order develops with increasing $\phi$, as shown in Fig.~\ref{fig:model}(c). To quantify this we analyse the standard functions used to follow orientational and translational order. For orientational order, we first determine the Voronoi tessellation in each configuration, and use the set of neighbours to evaluate the local degree of orientational order using
\begin{equation}
  \psi_6  ({\bf r}_i,t)=\frac{1}{\abs{\partial i}} \sum_{j\in\partial i} e^{6 i \theta_{ij}},
\end{equation}
where $\partial i$ is the set of neighbours of particle $i$ and $\theta_{ij}$ is the angle formed by ${\bf r}_i - {\bf r}_j$ and a reference axis. The global bond-orientational order parameter is then
\begin{equation}
    \Psi_6(t)=\frac{1}{N} \abs{\sum_{i} \psi_6( {\bf r}_i,t)}, 
\end{equation}
while the corresponding spatial correlations are measured using
\beq
g_6(r) = \frac{1}{\rho N} \expval{\sum_{i\neq j}\psi_6( {\bf r}_i)\psi_6^* ({\bf r}_j)\,\delta\qty(|{\bf r}-{\bf r}_j+{\bf r}_i|)} .
\eeq
For positional order we analyse both the pair correlation function 
\beq
g(r)=\frac{1}{\rho N}\expval{\sum_{i\neq j}\delta  (|{\bf r}-{\bf r}_j+{\bf r}_i|)}, 
\eeq
and the static structure factor $S(q)$ in Fourier space. 

In Fig.~\ref{fig:ordering}(a) we show the evolution of $\expval{\Psi_6}$ for several system sizes across $\phi_c$. In the absorbing phase, $\expval{\Psi_6}$ is small because contributions to the phase of $\Psi_6$ from different domains of the polycrystal cancel each other, and we expect $\expval{\Psi_6}=0$ in the thermodynamic limit. For $\phi>\phi_c$ the system has time to anneal all topological defects, $\expval{\Psi_6}$ is close to unity and only weakly decreases with $\phi$ as the activity $f_\infty$ increases and more particles perform a displacement at each time step. 

The color code in Fig.~\ref{fig:model}(c) describes orientational order which appears homogeneous on the scale of the polycrystalline grains. In Fig.~\ref{fig:ordering}(b) we show that translational order is also correlated over the same length, as both $g(r)$ and $g_6(r)$ are shown to decay as $\sim \exp(-r/\xi)$ for large $r$ with a similar lengthscale $\xi$. We show in Fig.~\ref{fig:ordering}(c) that $\xi(\phi)$ grows rapidly as $\phi_c$ is approached from below, and the evolution of $\xi$ is consistent with $\xi \sim (\phi - \phi_c)^{-\nu_{\perp}}$ with $\nu_\perp = 0.8$ the CDP value. This agreement suggests that structural ordering is governed by the physics of the absorbing phase transition.

Crystals do not exist at equilibrium in two dimensions because long-ranged translational order is destroyed by phonons~\cite{mermin1968crystalline}, and only quasi long-range order can survive~\cite{halperin1978theory,kosterlitz1973ordering,young1979melting}.  It is typically difficult to numerically distinguish these two scenarios, as finite size effects can be strong~\cite{bernard2011two}. In the active phase above $\phi_c$ we find that the system spontaneously orders from a random initial conditions and a similar steady state is reached when particles are first initialised on the sites of a perfectly ordered lattice. At steady state, a finite fraction $f_\infty$ of the particles performs a small jump at each step. Although reminiscent of Brownian motion, we now demonstrate that, differently from thermal fluctuations, this finite activity does not destroy long-ranged translational order.  

A strong quantitative indication stems from the mean-squared displacement (MSD)
\beq
\Delta^2(t) = \frac{1}{N} \sum_{i} \expval{ | {\bf u}_i(t)|^2 },
\label{eq:MSD}
\eeq
where ${\bf  u}_i(t)={\bf r}_i(t)-{\bf  r}_i(0)$. In Fig.~\ref{fig:activecrystal}(a) we report the MSD measured in the active phase for different system sizes and densities. For any $\phi > \phi_c$ the MSD reaches a well-defined plateau at long times, which is independent of $N$. This implies that particles remain close to their initial positions and that the crystalline order is stable. In equilibrium, time-averaged particles positions ${\bf R}_i$ are not defined and the long-time plateau of the MSD exists but its value grows logarithmically with system size, as a result of equipartition of energy. To see this, let us introduce the Fourier transform ${\bf u}_{\bf q}(t) = \sum_i {\bf u}_i \exp( i {\bf q} \cdot {\bf R}_i )$ and the corresponding structure factor for displacements,
$S_{u} ({\bf q})= \frac{1}{N} \langle {\bf u}_{\bf q}  \cdot {\bf u}_{-\bf q}   \rangle$.
Using a harmonic expansion of energy fluctuations in an equilibrium crystal~\cite{ashcroft2022solid}, each Fourier mode becomes excited by an amount $\langle | {\bf u}_{\bf q} |^2 \rangle \sim k_B T / q^2$ at low $q$ and temperature $T$. Together with the expression of the MSD plateau
\beq
\Delta^2(t \to \infty) \sim 
\int_{2\pi / L}^{\Lambda}  S_{u} (q)  q^{d-1} \de q,
\label{eq:MSD2}
\eeq
where $\Lambda \sim 1/a$ with $a$ the lattice spacing, one concludes that the integral in Eq.~(\ref{eq:MSD2}) diverges logarithmically with $L$ when $d=2$ and remains finite for $d>2$. A sufficient condition for the MSD to remain finite is to have $S_u(q)$ diverging slower than $q^{-2}$ so that the integral in Eq.~(\ref{eq:MSD2}) converges. This is equivalent to demanding that energy is not distributed in Fourier modes as imposed by equipartition of energy. We confirm in Fig.~\ref{fig:activecrystal}(b) that the measured $S_{u}(q)$ converges to a finite value as $q \to 0$, deviating strongly from a $1/q^2$ divergence. This finding rationalises the finite MSD reported in Fig.~\ref{fig:activecrystal}(a). The strong deviation from $q^{-2}$ in Fig.~\ref{fig:activecrystal}(b) implies a lack of equipartion of energy with low-$q$ displacement modes being strongly suppressed by the local dynamic rules. As a results, phonons are so strongly suppressed that they do not destroy long-range translational order: the breakdown of Mermin-Wagner theorem is thereby established.

A final confirmation that the active phase corresponds to two-dimensional nonequilibrium crystals is given by the static structure factor $S(q)$ shown in Fig.~\ref{fig:activecrystal}(c). We observe sharp Bragg peaks at discrete $q$ values with an amplitude that scales with $N$, resulting from the periodicity of the ordered lattice. We also observe a diffusive background at a much lower, $N$-independent, amplitude. For equilibrium crystals (in $d>2$), thermal fluctuations are responsible for this background, and in particular lead to a finite limit $S(q \to 0) \propto k_B T$ with a prefactor set by the elastic constants of the crystal~\cite{chaikin1995principles}. In our active phase, in contrast, the diffuse background has a peculiar behaviour at low $q$, $S(q) \sim q^2$, so that in particular $S(q \to 0) = 0$. A similar behaviour was reported for disordered diffusing states at lower density~\cite{hexner2017noise}. This suppression of fluctuations at low $q$ is required for long-range translational order to be sustainable in two dimensions. Put differently, crystals in $d=2$ (but not $d>2$) are necessarily hyperuniform~\cite{torquato2018hyperuniform}, whereas equilibrium thermal fluctuations in systems of finite compressibility destroy hyperuniformity (for any $d$)~\cite{ikeda2015thermal,kim2018effect}. 

Life out of equilibrium is typically richer than in equilibrium, as virtually any general theorem can be violated, paving the way to physical behaviours that have no equilibrium counterpart. Random organisation models are simple microscopic models where local breaking of detailed balance leads to rich physics at large scale. We showed that they provide a set of microscopic rules whereby two dimensional crystals with genuine long-range order become stable. Given the recent explosion of experimental realisations of two-dimensional nonequilibrium particle systems~\cite{marchetti2013hydrodynamics,zahn1999two,thorneywork2017two,reis2006crystallisation,olafsen2005twodimensional,komatsu2015roles,digregorio2018full,caprini2020hidden,briand2016crystallisation,gowrishankar2016nonequilibrium,ghosh2022coupled}, we encourage experimental studies of their properties in a parameter space where they can order.  

\acknowledgments

We thank J.-L. Barrat, D. Frenkel, and S. Ramaswamy for useful interactions. This project received funding from the Simons Foundation (\#454933, LB), a Visiting Professorship from the Leverhulme Trust (VP1-2019-029, LB), and by the European Research Council under the Horizon 2020 Programme, ERC grant agreement number 740269 (MEC). 

\bibliography{main.bib}

\end{document}